%% file: pyrite_261103-6.tex
\documentclass[%
reprint,
superscriptaddress,
amsmath,amssymb,
prb,
longbibliography,
]{revtex4-2}

\usepackage{dcolumn}
\input{revpackage.tex}

\begin{document}

\title{Electric toroidal octupolar symmetry in pyrite FeS$_2$ probed by Raman optical activity}

\author{Yuki~Suganuma}
\affiliation{%
Department of Physics, Institute of Science Tokyo, Tokyo 152-8551, Japan%
}%
\author{Gakuto~Kusuno}
\affiliation{%
Department of Physics, Institute of Science Tokyo, Tokyo 152-8551, Japan%
}%
\author{Hikaru~Watanabe}
\affiliation{%
Graduate School of Engineering, Hokkaido University, Sapporo 060-8628, Japan%
}%
\affiliation{%
Department of Physics, The University of Tokyo, Tokyo 113-0033, Japan%
}%
\author{Rikuto~Oiwa}
\affiliation{%
Graduate School of Science, Hokkaido University, Sapporo 060-0810, Japan%
}%
\author{Hitoshi~Mori}
\affiliation{%
Institute for Materials Research, Tohoku University, Sendai 980-8577, Japan%
}
\author{Ryotaro~Arita}
\affiliation{%
Department of Physics, The University of Tokyo, Tokyo 113-0033, Japan%
}%
\affiliation{Center for Emergent Matter Science, RIKEN, Wako 351-0198, Japan}
\author{Takuya~Satoh}
\email[]{satoh@phys.sci.isct.ac.jp}
\affiliation{%
Department of Physics, Institute of Science Tokyo, Tokyo 152-8551, Japan%
}%
\affiliation{%
Quantum Research Center for Chirality, Institute for Molecular Science, Okazaki 444-8585, Japan%
}%

\date{\today}

\begin{abstract}
In crystals with degenerate phonons, Raman optical activity (ROA), the difference in Raman
scattering intensity between circular polarization channels, can arise when the scattering
distinguishes components carrying opposite crystal angular momenta. In crystals it has long been associated with structural chirality and more recently with ferroaxial order, which are the electric toroidal multipoles of rank zero and rank one, but whether it can probe higher ranks has remained unknown. Here we report ROA in pyrite FeS$_2$, a centrosymmetric and nonmagnetic crystal whose $T_h$ symmetry forbids the electric toroidal multipoles of ranks zero, one, and two, so that the lowest-rank one it admits is the octupole of rank three. Using circularly polarized Raman scattering in a backscattering geometry, we
observe a pronounced circular intensity difference exclusively for the doubly degenerate $\Eg$
phonon, reaching about $85\%$ of the mean intensity at an excitation wavelength of 633 nm.
The difference reverses sign reproducibly between neighboring \{111\} facets and is strongly
suppressed on \{001\}. Group theory shows that this difference transforms as the antisymmetric part of the
self-product of the mode representation, which contains the octupolar $A_{2g}$ representation
for the $\Eg$ doublet but not for the $\Tg$ triplets and that it must vanish for
propagation along $\langle001\rangle$.
First-principles resonant Raman calculations, evaluated at the three experimental photon
energies within the electric-dipole approximation, reproduce the $\Eg$ selectivity and
the reversal of sign with excitation energy.
These results extend ROA from rank-one to rank-three axial
multipolar symmetry and establish it as a symmetry-selective, facet-resolved optical probe of
symmetries that produce neither macroscopic polarization nor magnetization.
\end{abstract}

\maketitle


\section{Introduction}

Multipolar degrees of freedom provide a unified framework for describing complex electronic and structural states in solids beyond conventional charge and spin dipoles. Multipoles are classified according to their transformation properties under spatial inversion and time-reversal symmetries into four fundamental categories: electric, magnetic, magnetic toroidal, and electric toroidal multipoles \cite{spaldin2008toroidal,hayami2016emergent,suzuki2017cluster,hayami2018microscopic,suzuki2019multipole,kusunose2020complete}. In particular, high-rank multipolar order can exist without inducing macroscopic polarization or magnetization, which makes it difficult to detect experimentally~\cite{pourovskii2025hidden}.

Experimental studies of multipolar order have progressed using techniques such as resonant x-ray and neutron scattering \cite{paixao2002triple,kusunose2005evidence,santini2009multipolar}, which provide relatively direct probes of multipolar degrees of freedom, and through bulk thermodynamic measurements interpreted with symmetry analysis \cite{matsuoka2006possible}. Where the multipole is magnetic, a tabletop route is also available. The cluster octupole of Mn$_3$Sn \cite{suzuki2017cluster} yields a large anomalous Hall \cite{nakatsuji2015large} and anomalous Nernst \cite{ikhlas2017large} response although the magnetization is vanishingly small. Where it is not, detection remains demanding and has relied on large-scale facilities and sophisticated symmetry analysis \cite{santini2009multipolar,bhowal2021revealing}.

Experimental knowledge of nonmagnetic octupolar order, including electric and electric toroidal octupoles, remains especially limited. Being even under time reversal, these multipoles lack the thermodynamic signatures, such as magnetization, through which time-reversal-odd multipoles are conventionally detected. Electric toroidal octupoles are in addition even under spatial inversion, so they produce no electric polarization either.
Identifying probes that are selectively sensitive to such high-rank axial multipolar symmetries thus remains an important open problem.

Recently, optical probes sensitive to symmetry breaking, such as second-harmonic generation, have emerged as powerful tools for accessing hidden electronic and structural orders, including ferroaxial order \cite{jin2020observation}.
In particular, Raman optical activity (ROA) is defined as the difference in Raman scattering intensity between left- and right-circularly polarized light \cite{barron1973raman}. In circularly polarized Raman scattering, a doubly degenerate phonon is resolved into components carrying opposite crystal angular momenta (CAM, often referred to as pseudoangular momenta~\cite{streib2021difference}) $m=\pm1$ \cite{zhang2014angular,zhang2015chiral,ishito2023a}. Throughout this work $m$ is defined by $C_3^{+}\bm{u}=e^{-2\pi i m/3}\bm{u}$, the sign convention of Refs.~\cite{zhang2015chiral,ishito2023a}, with $C_3^{+}$ the rotation $x\to y\to z\to x$ about $\langle111\rangle$. Structural chirality is known to distinguish phonon modes carrying opposite CAM and gives rise to chiral phonon phenomena in chiral crystals \cite{pine1969linear,ishito2023a,ishito2023b,Oishi2024,Kusuno2026a}.

More recently, ROA has been reported in several systems
with axial symmetry breaking \cite{lacinska2022raman,yang2022visualization,zhao2023spectroscopic,liu2023electrical}. In the chiral and polar NiCo$_2$TeO$_6$, the cross-circular intensity difference of the $E$ phonons was shown to be governed by the axial combination of structural chirality and electric polarization rather than by chirality alone, the two being separated by comparing twinned domains of opposite handedness~\cite{martinez2025ferroaxial}. A direct connection
between ROA and axial multipolar order was established for ferroaxial order in NiTiO$_3$ \cite{kusuno2026raman}. In ferroaxial crystals, this ROA originates from a rank-one electric toroidal dipole~\cite{kusuno2026raman}, which serves as the order parameter of ferroaxial transitions \cite{hlinka2016symmetry,bhowal2024electric}. The same axial moment produces a circular dichroism of the coupled phonons in resonant inelastic x-ray scattering as well~\cite{huang2025non}. These developments raise the question of whether the same principle extends to higher-rank axial multipoles. Can ROA detect a rank-three electric toroidal octupole, an axial symmetry that carries no dipolar component and produces neither macroscopic polarization nor magnetization?

Here we address this question by investigating phonon excitations in pyrite FeS$_2$, an earth-abundant and nontoxic semiconductor that has been pursued as a low-cost photovoltaic material \cite{uchiyama2021optical}.
Pyrite crystallizes in the centrosymmetric $T_h$ ($m\bar{3}$) point group, where, despite the absence of magnetism or structural chirality, an electric toroidal octupole is symmetry allowed \cite{hayami2018classification,yatsushiro2021multipole}. Of the seven components of a rank-three axial multipole, $T_h$ leaves only one invariant, and it has the $xyz$ angular form, $G_{xyz}$.
This octupolar symmetry is a property of the known $Pa\bar{3}$ structure, not an ordered phase reached through a phase transition. What has been missing is an observable that couples to it. The same $xyz$-type octupole is allowed in the chiral class $23$, reached by replacing the S$_2$ dimer with an ordered pair of two different anions as in ullmannite, a reduction that the structure determination of pyrite itself notes gives virtually the same atomic arrangement~\cite{Brostigen1969}.
Recent theory \cite{watanabe2025dual} predicts that cross-circular ROA provides a symmetry-sensitive probe of axial multipolar symmetry, including electric toroidal octupoles, arising from mirror-symmetry breaking.

\begin{figure*}
\centering
\includegraphics[width=\textwidth]{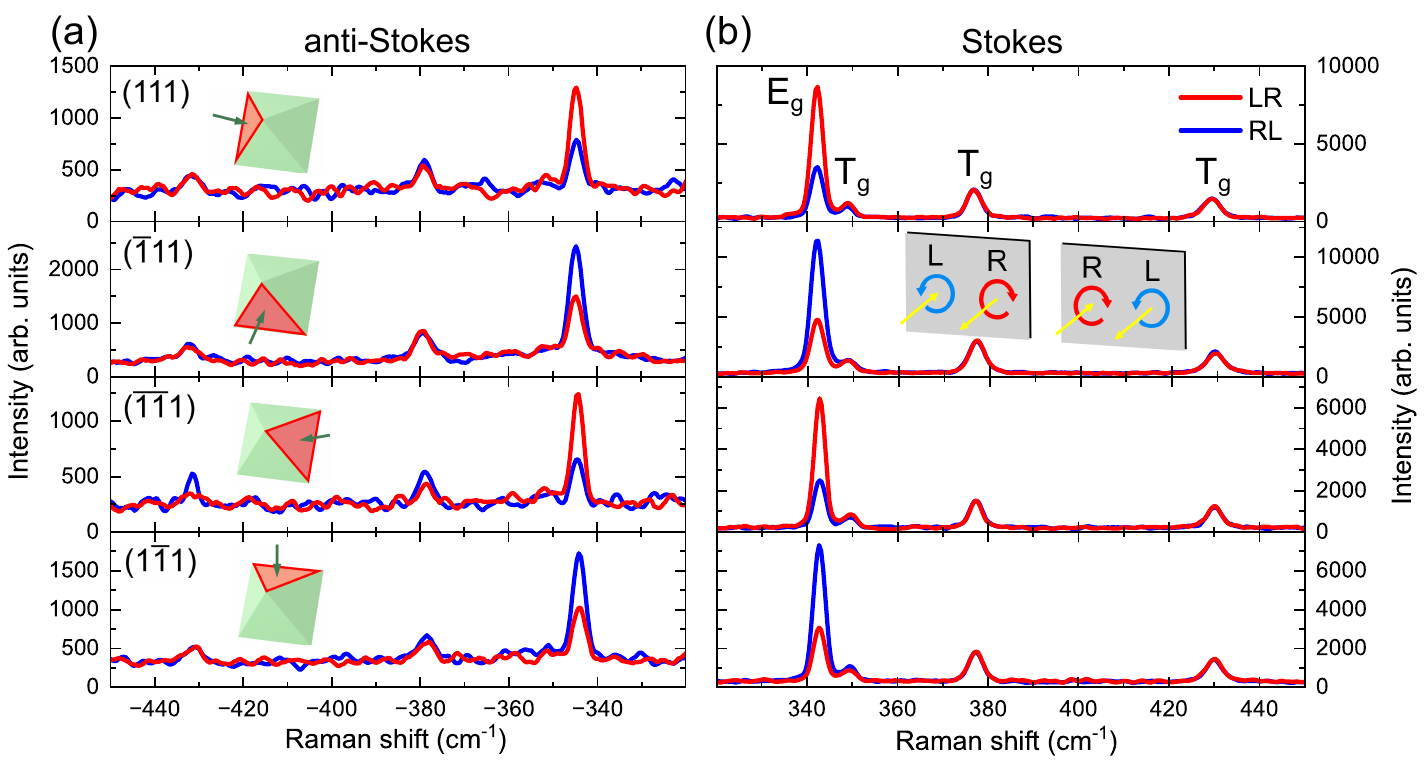}
\caption{
(a) Anti-Stokes and (b) Stokes Raman spectra measured on \{111\} surfaces at an excitation wavelength of 633 nm in the cross-circular polarization
configurations (LR in red; RL in blue).
Insets in (a): the octahedral pyrite crystal, with the measured facet highlighted in red and the arrow indicating the propagation direction of the
incident light, i.e., the inward normal of that facet (backscattering geometry).
Inset in (b): schematic illustrations of the cross-circular polarization configurations (LR and RL) in a backscattering setup.
The labels $(111)$, $(\bar{1}11)$, $(\bar{1}\bar{1}1)$, and $(1\bar{1}1)$ indicate the crystal orientation.
}
\label{111abcd633}
\end{figure*}

\section{Methods}

\subsection{Samples and circularly polarized Raman spectroscopy}

We performed circularly polarized Raman scattering in a backscattering geometry on octahedral and cubic pyrite single crystals obtained from natural mineral specimens. Measurements were carried out at 295 K using a home-built Raman spectrometer with excitation wavelengths of 532, 633, and 785 nm.
The cross-circular polarization configurations (LR and RL), where the first and second letters denote the polarization of the incident and scattered light, respectively, are illustrated in the inset of Fig. \subref{111abcd633}{b}. The four faces of the octahedral crystal are labeled as $(111)$, $(\bar{1}11)$, $(\bar{1}\bar{1}1)$, and $(1\bar{1}1)$ [insets of Fig. \subref{111abcd633}{a}].
The two cross-circular configurations are interchanged by rotating a single quarter-wave
plate, which the incident and the scattered beam both traverse, by $90^{\circ}$. Every
other element of the setup, and the alignment, is left untouched. Peak positions and
intensities were obtained from Voigt fits. The optical layout, the instruments, and the
measurement conditions are given in the Supplemental Material \cite{SM} (Sec. S1).

\subsection{First-principles calculations}
\label{sec:calc}

To verify the symmetry-based interpretation of the multipolar phonon Raman response,
the phonon spectrum, the mode symmetries, and the resonant Raman intensities were all
obtained from the same density-functional calculation. The lattice dynamics was treated
within density-functional perturbation theory~\cite{Baroni2001} as implemented in
\textsc{quantum espresso}~\cite{Giannozzi2009}, using norm-conserving pseudopotentials
generated with the optimized norm-conserving Vanderbilt scheme~\cite{Hamann2013} and the
Perdew--Burke--Ernzerhof exchange-correlation functional~\cite{Perdew1996}, with the
structure relaxed within the same functional. The Raman intensities were evaluated from
Wannier-interpolated electronic states using \textsc{wannier90}~\cite{Pizzi2020},
\textsc{epw}~\cite{Lee2023}, and \textsc{qr2code}~\cite{Huang2026}.
For phonon mode $\nu$ and incident photon energy $E_{\mathrm{L}}$, the calculated
polarization-resolved Raman amplitude is
\begin{equation*}
M_{\nu}(E_{\mathrm{L}}) = \mathbf{e}_{s}^{\dagger}\,\hat{\mathcal{R}}_{\nu}(E_{\mathrm{L}})\,\mathbf{e}_{i},
\qquad
I_{\nu}(E_{\mathrm{L}}) \propto |M_{\nu}(E_{\mathrm{L}})|^{2},
\end{equation*}
where $\hat{\mathcal{R}}_{\nu}(E_{\mathrm{L}})$ is the complex resonant Raman tensor for the
one-phonon process, constructed from electron-photon and electron-phonon matrix elements.
For the two $\Eg$ components, for which we write $\nu = \pm$, this tensor is the first-principles counterpart of the symmetry-adapted Raman tensor $\hat{\alpha}$ of the Appendix, with $\hat{\mathcal{R}}_{\pm} = \Phi_{\pm}\hat{\chi}^{(\pm)}$.
The cross-circular response of this tensor for axial multipolar order was formulated in
Ref.~\cite{watanabe2025dual}, where it was evaluated for pyrite as a function of photon energy.
Here the same calculation is carried out at the three photon energies used in the experiment and
for all Raman-active modes.
In the present calculation the Raman tensor is evaluated for the channel in which the
photoexcited electron is scattered by the phonon within the conduction band and the
calculations were performed within the electric-dipole approximation. Further details,
including the relaxed structure, the computational parameters, the explicit form of
$\hat{\mathcal{R}}_{\nu}$, and the definition of the mechanical angular momentum, are given in
the Supplemental Material \cite{SM} (Sec. S2).

\section{Results}

\subsection{Raman spectra and mode assignments}

Figures \subref{111abcd633}{a} and \subref{111abcd633}{b} show anti-Stokes and Stokes Raman spectra, respectively, measured on four adjacent \{111\} faces of an octahedral pyrite crystal under cross-circular polarization configurations with an excitation wavelength of 633 nm.
Because the photon momentum in Raman scattering is negligible compared with the Brillouin-zone size, the measured spectra probe phonons near the $\Gamma$ point. The calculated phonon dispersion of pyrite FeS$_2$, including the Raman-active $\Ag$, $\Eg$, and $T_g$ modes \cite{verble1969infrared,ushioda1972raman,vogt1983complete}, is presented in Sec.~\ref{sec:disp} and is consistent with the experimentally observed Raman peak positions.

A pronounced circular intensity difference between the LR and RL configurations is observed exclusively for the doubly degenerate $\Eg$ phonon mode, while the $\Tg$ modes show no detectable circular intensity difference within the experimental noise level.
Measurements under parallel-circular polarization configurations (LL and RR) yielded no significant circular intensity difference for any of the phonon modes [Supplemental Material \cite{SM} (Sec. S4)].

For the pyrite structure containing four FeS$_2$ units per primitive cell, the $\Gamma$-point phonon modes consist of 36 vibrational modes in total.
Among them, 12 modes are Raman active, corresponding to the irreducible representations
$\Ag + \Eg + 3\Tg$, as established by group-theoretical analysis of the pyrite lattice~\cite{verble1969infrared}.
Here, the $\Eg$ mode is doubly degenerate and each $\Tg$ mode is triply degenerate.
The complete zone-center decomposition is $\Gamma = \Ag + \Eg + 3\Tg + 2A_u + 2E_u + 6T_u$. Because pyrite is centrosymmetric the mutual exclusion rule applies~\cite{dresselhaus2008group}, so the even-parity modes listed above are Raman active while the odd-parity modes are either infrared active ($5T_u$, one $T_u$ being acoustic) or silent ($2A_u+2E_u$). The irreducible representation of each mode in Table~\ref{FeS2_phonon_frequencies} was assigned from the degeneracy and the symmetry characters of the computed eigenvectors, which separates the $\Eg$ doublet from the nearby silent $A_u$ singlet.

The calculated phonon frequencies at the $\Gamma$ point are summarized in Table~\ref{FeS2_phonon_frequencies}, together with the experimental values obtained in this work and those reported in Ref.~\cite{vogt1983complete}.
The overall agreement with experiment, as well as with the phonon dispersion measured by inelastic neutron scattering~\cite{buhrer1993lattice}, confirms the validity of the present calculations. The largest deviation is that of the $\Ag$ mode, which lies $5\%$ below the reported value and, in the calculation, below $\Tg^{(2)}$ rather than above it. The $\Ag$ mode is the pure S--S stretch and is therefore the most sensitive of the five to the calculated bond length, which exceeds the measured one by $1.06\%$ [Supplemental Material \cite{SM} (Sec. S2)].

\begin{table}[htbp]
\centering
\caption{Frequencies and symmetries of Raman-active phonon modes at the $\Gamma$ point in pyrite FeS$_2$. Experimental values from this work and Ref. \cite{vogt1983complete} are also provided for comparison. No experimental value is listed for the $\Ag$ mode because its Raman tensor is isotropic, so that the mode is strictly forbidden in the cross-circular configurations used here. It is observed in the parallel-circular configurations instead [Supplemental Material \cite{SM} (Sec. S4)].}
\setlength{\tabcolsep}{8pt}
\begin{tabular}{cccccc}
\hline
\hline
& $\Eg$ & $\Tg^{(1)}$ & $\Ag$ & $\Tg^{(2)}$ & $\Tg^{(3)}$ \\
\hline
Calc. (cm$^{-1}$) & 332 & 338 & 360 & 380 & 420 \\
Expt. (cm$^{-1}$) & 342 & 349 & - & 377 & 429  \\
Ref. \cite{vogt1983complete} (cm$^{-1}$) & 343 & 350 & 379 & 377 & 430 \\
\hline
\hline
\end{tabular}
\vspace{3ex}
\label{FeS2_phonon_frequencies}
\end{table}

\subsection{Facet dependence of the circular intensity difference}
\label{sec:facet}

A key experimental signature is the sign reversal of the circular intensity difference between neighboring \{111\} faces. 
As the measurement position moves sequentially across adjacent \{111\} faces, the sign alternates in a $+$ $-$ $+$ $-$ sequence. The four faces are crystallographically equivalent, being related to one another by the symmetry operations of $T_h$, so x-ray diffraction cannot tell them apart. What distinguishes them is the sign of the cross-circular response, which is odd under the improper operations that relate them. This alternating behavior is highly reproducible and is observed for multiple sets of adjacent \{111\} faces on another octahedral pyrite sample.

The same sign alternation is observed for both Stokes and anti-Stokes scattering. This distinguishes the present effect from magnetic order, since in magnon Raman scattering the Stokes and anti-Stokes components obey opposite circular selection rules~\cite{cenker2021direct} and more generally the relation between the Stokes and the anti-Stokes circular intensity difference is fixed by the antiunitary symmetry that the system retains~\cite{barron1982anti,watanabe2025symmetry}. The same criterion has been derived for the octupolar case~\cite{watanabe2025dual}. The octupole relevant here is $G_{xyz}$, identified in Sec.~\ref{sec:octupole}. It gives Stokes and anti-Stokes differences of the same sign if it is even under time reversal and of opposite signs if it is odd. The present measurements give the same sign for both, which is the time-reversal-even case.

The circular intensity difference is quantified by $g_{\mathrm{ROA}}=2(\ILR-\IRL)/(\ILR+\IRL)$, extracted from Voigt-function fits to the Raman peaks. The magnitude $|g_{\mathrm{ROA}}|$ reaches approximately 0.85 at 633 nm on the \{111\} surfaces (Table \ref{t1}). Switching the incident and scattered circular polarizations together, as in the LR and RL configurations used here, is the dual-circular-polarization form of ROA proposed for molecules~\cite{nafie1989dual} and first realized in backscattering by Che \textit{et al.}~\cite{che1991dual}. At 633~nm the $\Eg$-mode intensities scale linearly with the incident laser power over $0.5$--$5.0$~mW and $g_{\mathrm{ROA}}$ is independent of it, with no detectable contribution from local laser heating or nonlinear processes [Supplemental Material \cite{SM} (Sec. S6)]. The $5.5$~mW used for the 633-nm spectra lies just above that range, where the linear trend shows no sign of saturation.

\begin{table}
\centering
\caption{Values of $g_{\mathrm{ROA}}$ obtained by fitting the $\Eg$ mode intensities at different excitation wavelengths. The fitting uncertainty in $g_{\mathrm{ROA}}$ is below 0.04 for all data points.}\label{t1}
\begin{tabular}{cccccc}
\hline
\hline
\noalign{\vspace{2pt}}
Wavelength & ~$(111)$ & ~$(\bar111)$ & ~$(\bar1\bar11)$ & ~$(1\bar11)$ & ~\{001\} \\
\noalign{\vspace{1pt}}
\hline
\noalign{\vspace{1pt}}
532 nm (Stokes)      &~$+$0.04~&~$-$0.05~&~$+$0.07~&~$-$0.04~&    -     \\
\noalign{\vspace{1pt}}
633 nm (Stokes)      &~$+$0.85~&~$-$0.84~&~$+$0.86~&~$-$0.88~&~$+$0.03~\\
\noalign{\vspace{1pt}}
633 nm (anti-Stokes)&~$+$0.68~&~$-$0.74~&~$+$0.84~&~$-$0.66~&    -     \\
\noalign{\vspace{1pt}}
785 nm (Stokes)      &~$-$0.08~&~$+$0.13~&~$-$0.08~&~$+$0.07~&    -     \\
\noalign{\vspace{1pt}}
\hline
\hline
\end{tabular}
\end{table}

\subsection{Suppression on the \{001\} surface}

\begin{figure}[htbp]
\centering
\includegraphics[width=0.7\hsize]{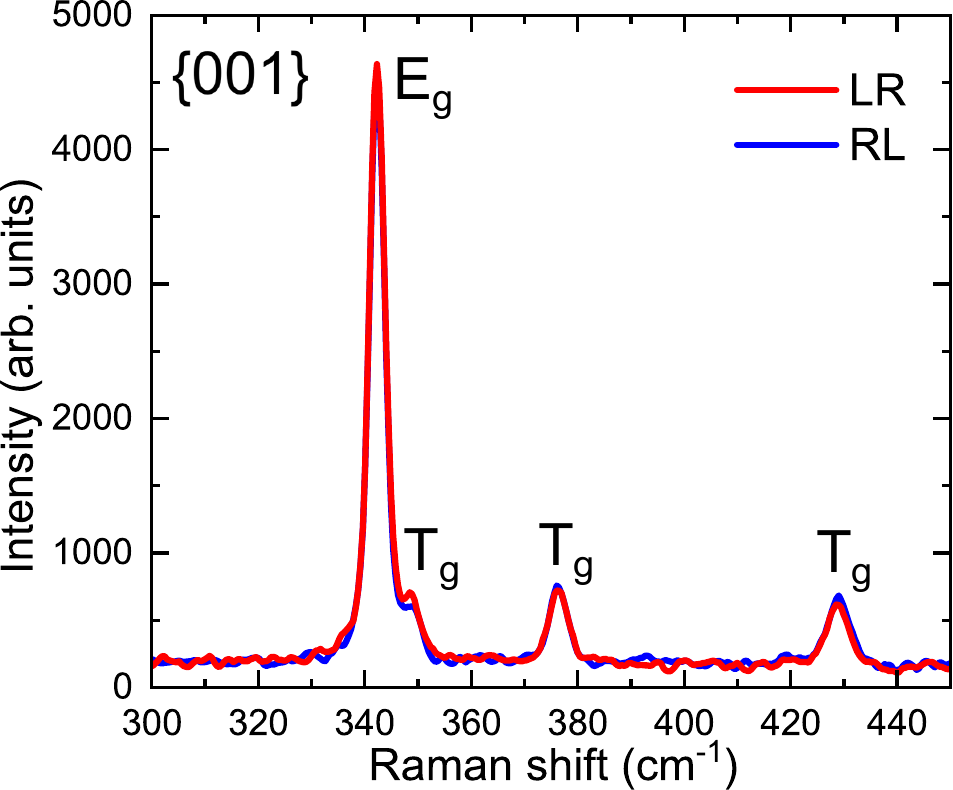}
\caption{Stokes Raman spectra measured on the \{001\} surface at 633~nm in the cross-circular (LR and RL) configurations. In contrast to the \{111\} surfaces (Fig.~\ref{111abcd633}), the circular intensity difference is strongly suppressed.}
\label{fig:s001}
\end{figure}

Figure~\ref{fig:s001} shows the Stokes spectra measured on the \{001\} surface. The circular intensity difference is strongly suppressed there, in contrast to the \{111\} facets. Together with the facet alternation of Sec.~\ref{sec:facet}, this establishes a facet-orientation-dependent ROA that directly reflects the axial multipolar symmetry of the crystal. The symmetry origin of this suppression is given in Sec.~\ref{sec:octupole}.

\subsection{Excitation-wavelength dependence}

\begin{figure}[htbp]
\centering
\includegraphics[width=\hsize]{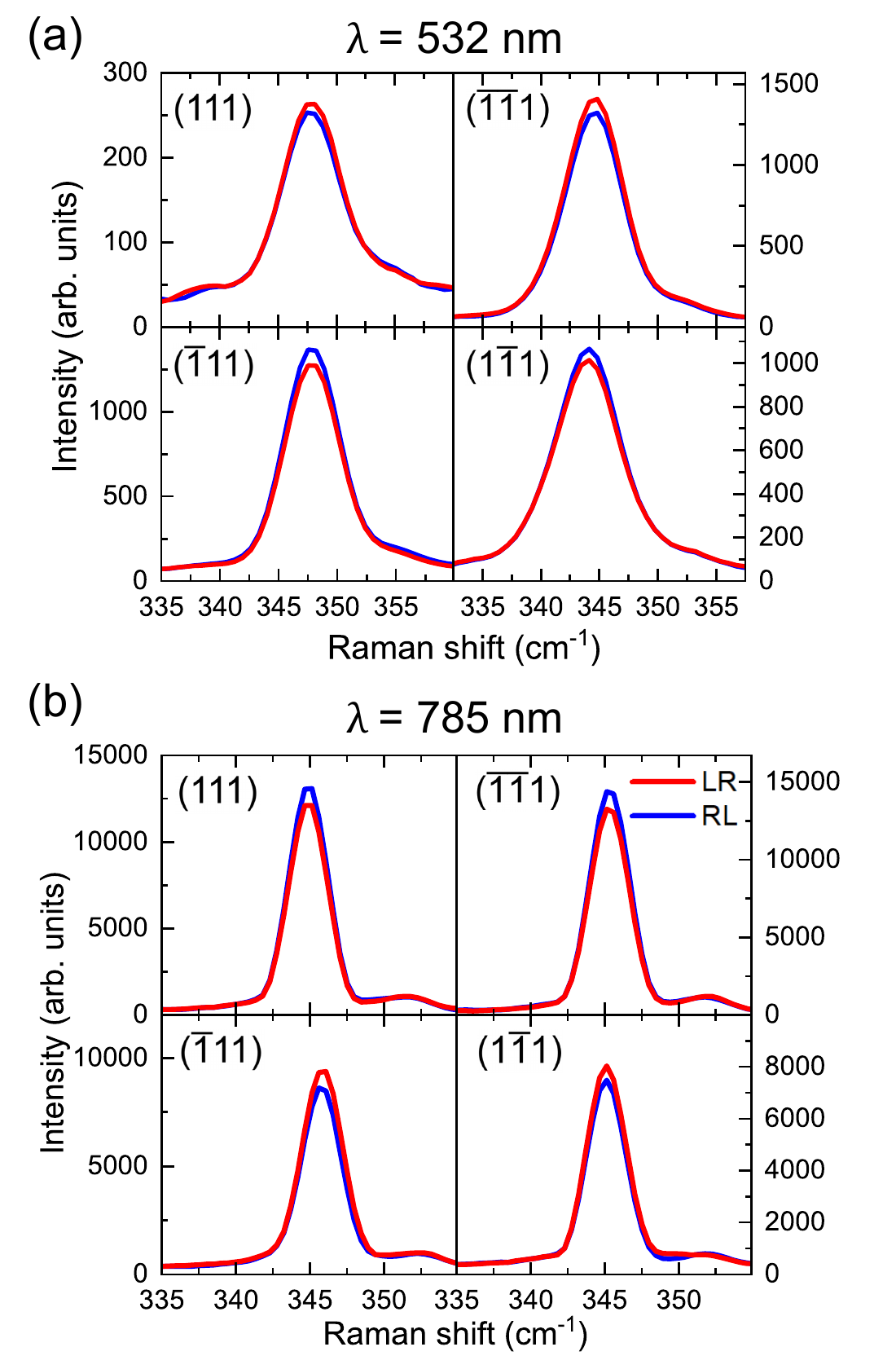}
\caption{Stokes Raman spectra measured on the \{111\} surfaces at excitation wavelengths of (a) 532~nm and (b) 785~nm in the cross-circular configurations, zoomed in on the $\Eg$ phonon mode. The 633~nm data are shown in Fig.~\ref{111abcd633}.}
\label{fig:swav}
\end{figure}

Figure~\ref{fig:swav} shows the cross-circular Raman spectra measured on the four \{111\} facets at excitation wavelengths of 532 and 785~nm, complementing the 633~nm data of Fig.~\ref{111abcd633}. The circular intensity difference remains confined to the $\Eg$ mode at all three wavelengths. At 532~nm the signal is an order of magnitude weaker than at 633~nm. Although the individual $|g_{\mathrm{ROA}}|$ values are comparable to the fitting uncertainty, they reproduce the same $+$ $-$ $+$ $-$ facet alternation and phase as at 633~nm (Table~\ref{t1}). This consistent, symmetry-locked pattern supports a small but finite signal. At 785~nm the signal is likewise weak but reverses its sign relative to 633~nm.

\subsection{Calculated phonon dispersion and crystal angular momentum}
\label{sec:disp}

\begin{figure}[htbp]
\centering
\includegraphics[width=\hsize]{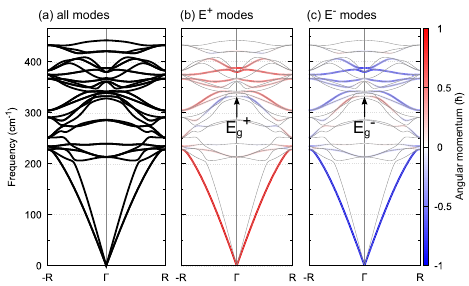}
\caption{
(a) Calculated phonon dispersion of pyrite FeS$_2$ along $\langle111\rangle$, from $-$R through $\Gamma$ to R, obtained using density-functional perturbation theory.
(b), (c) Angular momentum $\mathbf{L}$ projected along the $\langle111\rangle$ axis for the two orthogonal components, (b) $E^{+}$ and (c) $E^{-}$. In the calculations, all degenerate phonon modes are decomposed into orthogonal components. The full results are shown and we focus on the $\Eg$ mode, which is relevant to the Raman optical activity.
}
\label{FeS2_band}
\end{figure}

Figure~\subref{FeS2_band}{a} shows the calculated phonon dispersion along $\langle111\rangle$.
At the $\Gamma$ point, the phonon modes belong to the $T_h$ point group.
The doubly degenerate $\Eg$ phonon at the $\Gamma$ point decomposes into the two orthogonal components $\Egn{+}$ and $\Egn{-}$, labeled by their CAM $m=\pm1$, and the doublet remains degenerate along $\Gamma$--R, where the little group is $C_3$ and the components are written $E^{+}$ and $E^{-}$.
The mechanical angular momentum $\mathbf{L}$ was evaluated for each component of the degenerate modes and projected along the $\langle111\rangle$ axis [Figs.~\subref{FeS2_band}{b} and \subref{FeS2_band}{c}].
For the $\Eg$ components it vanishes in the small-wave-vector limit.
The vanishing is not a numerical coincidence but a consequence of the cubic symmetry. $\mathbf{L}$ is an axial vector and therefore transforms as $\Tg$ in $T_h$, whereas the quantities that can be formed within the doublet are limited to $\Eg\otimes\Eg = 2A_g\oplus\Eg$ in $T_h$, which contains no $\Tg$~\cite{dresselhaus2008group}. The two $A_g$ terms are the descendants of $A_{1g}$ and $A_{2g}$ of the parent $O_h$ group, in which the same product reads $A_{1g}\oplus A_{2g}\oplus\Eg$ (Sec.~\ref{sec:rules}). The angular momentum of either $\Eg$ component is thus exactly zero at $\Gamma$. The individual atoms may move on ellipses, but the sum over the cell cancels identically. The same counting does not forbid it for the triplets, whose self-product has an antisymmetric part that is itself the axial-vector representation $\Tg$. The finite values carried by those branches at $\Gamma$ in Figs.~\ref{FeS2_band}(b) and \ref{FeS2_band}(c) are of that kind and it is only the doublet that is required to vanish. The vanishing therefore holds only in the cubic groups, where the axial vector is three-dimensional. In a uniaxial ferroaxial crystal such as NiTiO$_3$ the corresponding component of $\mathbf{L}$ belongs to the identity representation, so it is not required to vanish \cite{kusuno2026raman}.

\subsection{Calculated circular Raman spectra}
\label{sec:calcspec}

\begin{figure}[!t]
\hspace{-19pt}\includegraphics[width=1.0\hsize]{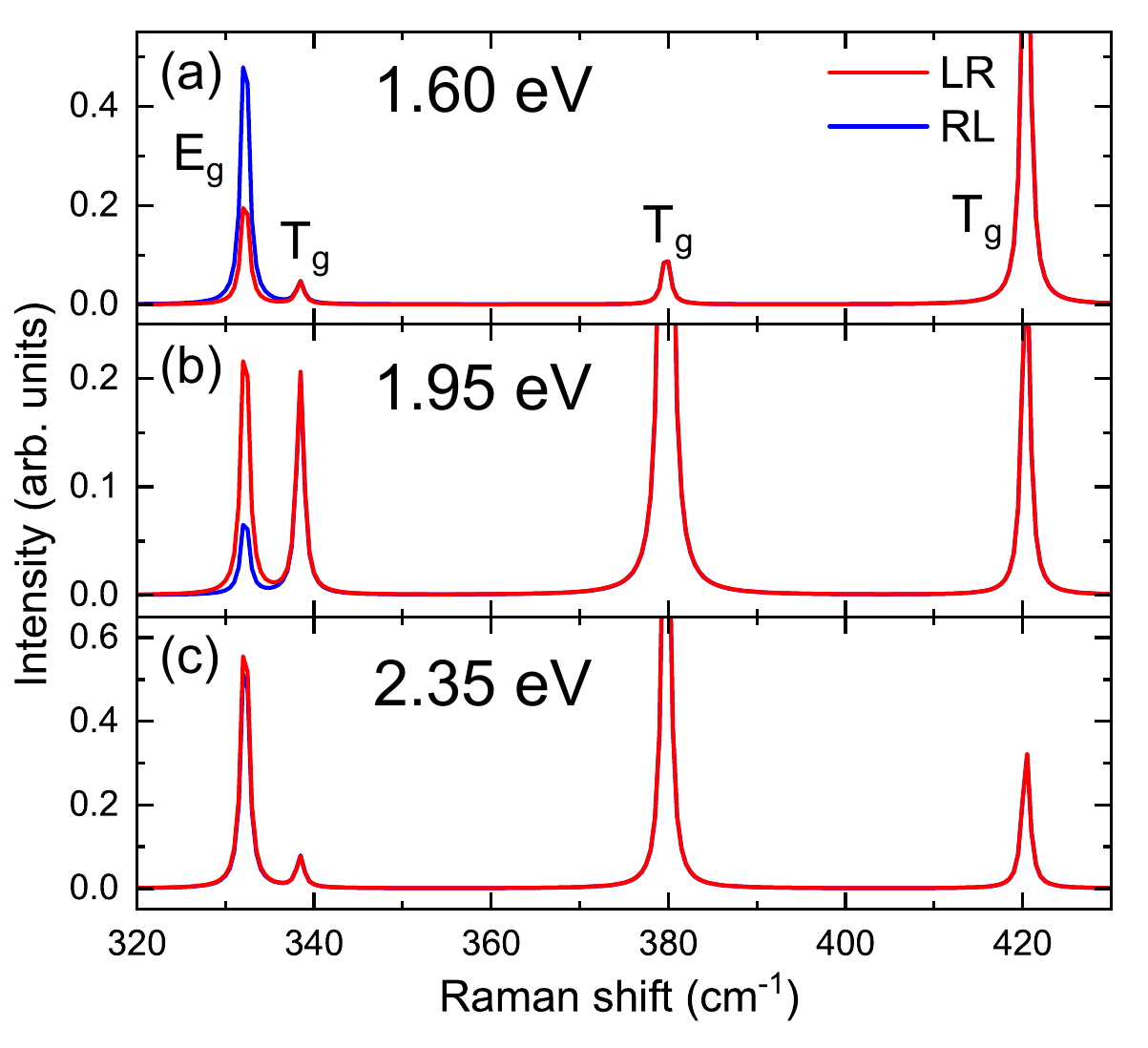}
\caption{
First-principles Raman spectra of pyrite FeS$_2$ calculated for the cross-circular polarization configurations (LR in red; RL in blue) on a representative \{111\} surface, at excitation energies of (a) 1.60, (b) 1.95, and (c) 2.35~eV, chosen to correspond to the
experimental wavelengths of 785, 633, and 532~nm.
}
\label{DFT_spectra}
\end{figure}

Figures~\subref{DFT_spectra}{a}, \subref{DFT_spectra}{b}, and \subref{DFT_spectra}{c} show the calculated Raman spectra for the cross-circular configurations on a representative \{111\} surface at excitation energies of 1.60, 1.95, and 2.35~eV.
The calculations reproduce the key experimental features. The circular intensity difference appears only for the $\Eg$ phonon mode and vanishes for the $T_g$ modes and the sign of the $\Eg$-mode circular intensity difference reverses between 1.95 and 1.60~eV, in agreement with the sign reversal observed experimentally between 633 and 785~nm (Table~\ref{t1}). At 2.35~eV, corresponding to 532~nm, the calculated $\Eg$-mode circular intensity difference is small, consistent with the weak signal measured at 532~nm (Table~\ref{t1}). At $1.60$~eV, by contrast, the calculated difference remains comparable to that at $1.95$~eV, whereas the measured $|g_{\mathrm{ROA}}|$ at 785~nm is an order of magnitude smaller than at 633~nm. This disagreement in magnitude is examined in Sec.~\ref{sec:resonance}.
We note that the calculated $T_g$-mode intensities and peak positions show some deviations from experiment, reflecting the sensitivity of the computed spectra to the underlying electronic structure of FeS$_2$. Nevertheless, the sign of the $\Eg$-mode circular intensity difference, which is the central observable here, is reproduced.

\section{Discussion}

To interpret these experimental observations, we now discuss the symmetry origin of the observed ROA.

\subsection{Circular-polarization selection rules}
\label{sec:rules}

The microscopic origin of the ROA can be discussed in terms of the coupling between phonons and the electronic structure mediating the Raman process \cite{watanabe2025dual}. 
The $\Eg$ phonon mode in pyrite FeS$_2$ is doubly degenerate and can be decomposed into two orthogonal multipolar vibrational modes, $\Egn{+}$ and $\Egn{-}$, that transform differently under mirror operations.
The atomic displacement patterns of these modes obtained from first-principles calculations are visualized in Supplemental Movie~1, with a representative snapshot shown in the Supplemental Material \cite{SM} (Sec.~S3). As the movie makes clear, the collective displacement pattern of each component involves all three crystallographic directions and cannot be confined to a single line or plane of the crystal. Each is a multipolar phonon carrying a definite CAM $m=\pm1$ along $\langle111\rangle$. The cyclic character of these modes lies in the order in which the distortion stretches and contracts along $[100]$, $[010]$, and $[001]$, the two components running through that cycle in opposite senses~\cite{watanabe2025dual}, and not in a circulation of the atoms. The phonon angular momentum $\mathbf{L}$ of the two $\Eg$ components ($\Egn{+}$ and $\Egn{-}$) vanishes at $\Gamma$ (Sec.~\ref{sec:disp}). It is $m=\pm1$, rather than any net rotation of the lattice, that the octupolar symmetry renders inequivalent.
Under cross-circular polarization, the LR and RL configurations selectively excite phonons with CAM $m=+1$ and $m=-1$, respectively \cite{ishito2023a}. In the presence of an electric toroidal octupolar anisotropy, the two CAM states ($m=\pm1$), which are otherwise related by mirror symmetry, become optically inequivalent. This inequivalence gives rise to a finite circular intensity difference in Raman scattering.

This microscopic picture can be expressed more formally using the complex Raman tensor of the $\Eg$ phonons. The LR and RL scattering geometries selectively couple to the complex components $\Phi_{+}$ and $\Phi_{-}$, respectively, which are the normal coordinates that create the $m=+1$ and $m=-1$ phonons (the Appendix). As a result, the Raman intensities for light incident along the $[111]$ direction can be written as
\begin{equation}
I^{[111]}_{\mathrm{LR}} = |\chi_{+}\Phi_{+}|^2,
\qquad
I^{[111]}_{\mathrm{RL}} = |\chi_{-}\Phi_{-}|^2.
\end{equation}
Here $\chi_{\pm}$ are the Raman susceptibilities corresponding to $\Phi_{\pm}$.
The difference between the LR and RL configurations therefore leads to
a circular intensity difference
\begin{equation}
U^{[111]}_{\mathrm{CC}}
=
\left|\chi_{+} \Phi_{+}\right|^2
-
\left|\chi_{-} \Phi_{-}\right|^2 .
\label{eq:UCC}
\end{equation}
The two $\Eg$ components are degenerate and are related to each other by time reversal, so they
are equally excited in thermal equilibrium and their amplitudes have equal magnitudes,
$|\Phi_{+}| = |\Phi_{-}|$~\cite{watanabe2025dual}. The amplitudes then cancel from Eq.~(\ref{eq:UCC}), leaving $U_{\mathrm{CC}} \propto |\chi_{+}|^2 - |\chi_{-}|^2$. The circular intensity difference therefore measures how unequally the two components scatter light, not how unequally they are populated.
Because the two $\Eg$ components are exchanged between neighboring \{111\} crystal faces by the $\{100\}$ mirror operations of $T_h$, the sign of $U_{\mathrm{CC}}$ reverses accordingly. The one-phonon scattering amplitude is linear in the phonon normal coordinates, so the intensity, being its square, is quadratic in them, and any intensity difference within a degenerate multiplet transforms in the direct product $\Gamma\otimes\Gamma$ of the mode representation $\Gamma$~\cite{dresselhaus2008group}. We use the parent $O_h$ group here because the octupolar channel is a distinct representation, $A_{2g}$, only in $O_h$. The cross-circular intensity difference is in addition antisymmetric under interchange of the two degenerate partners, so it transforms as the antisymmetric part of that product and the octupole can generate it only when the octupolar $A_{2g}$ representation appears there. For the $\Eg$ doublet, $\Eg\otimes\Eg=A_{1g}\oplus A_{2g}\oplus\Eg$ in $O_h$, whose antisymmetric part is the $A_{2g}$ representation of $G_{xyz}$. The octupole can therefore render the two $m=\pm1$ components optically inequivalent and produce a circular intensity difference. For the triply degenerate $T_g$ modes (Raman-active $T_{2g}$ in $O_h$), $T_{2g}\otimes T_{2g}=A_{1g}\oplus\Eg\oplus T_{1g}\oplus T_{2g}$ contains no $A_{2g}$, so no net circular intensity difference arises (Fig.~\ref{111abcd633}). Among the Raman-active phonons of pyrite, the selection rule therefore picks out only the doubly degenerate $\Eg$ mode. The same conclusion follows without reference to the parent group. In $T_h$ itself the antisymmetric part of $\Eg\otimes\Eg$ is the totally symmetric $A_g$, into which the $O_h$ octupolar $A_{2g}$ descends, whereas the antisymmetric part of $\Tg\otimes\Tg$ is $\Tg$ and contains no $A_g$. Carrying $m=\pm1$ components is therefore necessary but not sufficient. The $T_g$ triplet does decompose into $m=0,\pm1$ sectors under the threefold rotation about $\langle111\rangle$, yet the octupole cannot render its $m=\pm1$ sectors inequivalent because the $A_{2g}$ channel is absent from $T_{2g}\otimes T_{2g}$. A detailed derivation based on the complex Raman tensor of the $\Egn{+}$ and $\Egn{-}$ modes is given in the Appendix.

\subsection{Octupolar assignment and its real-space origin}
\label{sec:octupole}

In a backscattering geometry with cross-circular polarization, the circular intensity difference is odd under any mirror reflection whose mirror plane contains the light propagation direction [111] \cite{watanabe2025dual}. A finite ROA along $[111]$ therefore requires that the crystal retain no such mirror, which holds here because the planes that contain a $\langle111\rangle$ direction are of $\{110\}$ type and are absent in $T_h$, and the sign reversal between the mirror-related orientations $(111)$ and $(\bar111)$ follows from the $\{100\}$ mirrors that $T_h$ does retain.

\begin{figure}[!t]
\centering
\includegraphics[width=\linewidth]{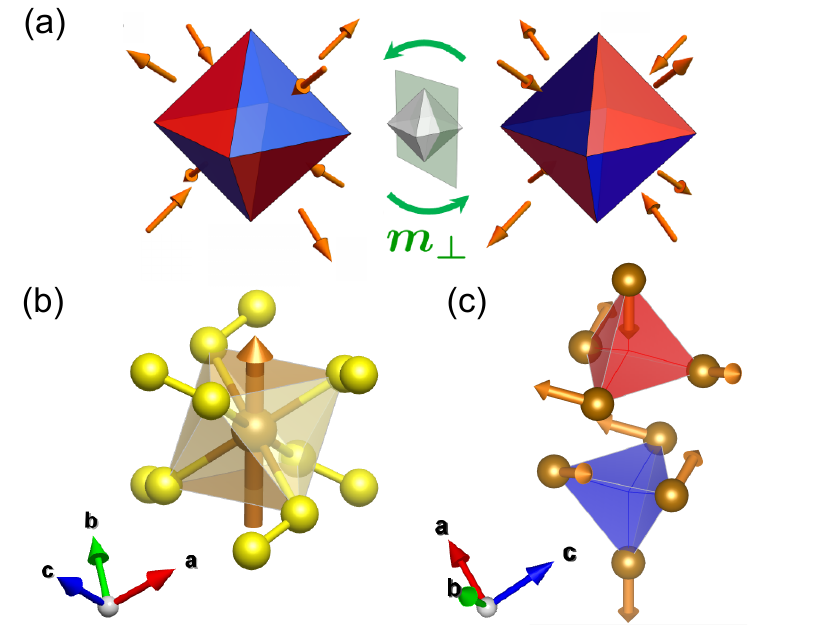}
\caption{
(a) Schematic of the $xyz$-type electric toroidal octupole $G_{xyz}$ and its transformation under a mirror reflection $m_{\perp}$ whose plane contains a $\langle 111 \rangle$ axis, after Ref.~\cite{watanabe2025dual}. The octahedron is a guide to the eye carrying the eight $\langle 111 \rangle$ directions. Each face is colored according to whether its axial vector (orange arrow) emerges from (red) or enters (blue) it, so the eight arrows represent four axial vectors. Under $m_{\perp}$, which is not a symmetry of $T_h$, red and blue are interchanged and the octupole changes sign, whereas the $\{100\}$ mirrors of $T_h$ leave it invariant.
(b), (c) Real-space origin, visualized with VESTA~\cite{momma2011vesta}. (b) A single FeS$_6$ cluster with the S$_2$ dimer partner of each S atom. The resulting ferroaxial moment (orange arrow) points along the local $\langle 111 \rangle$ axis. (c) The ferroaxial moments of the four Fe sublattices, viewed with a $\langle111\rangle$ axis vertical. Red and blue mark the two sublattice tetrahedra. Related by the nonsymmorphic $2_1$ screws of $Pa\bar{3}$, they form an all-in-all-out texture that realizes $G_{xyz}$, shown in (a).
}
\label{A2g}
\end{figure}

In the $T_h$ point group of pyrite FeS$_2$, the electric toroidal monopole $G_0$ associated with structural chirality, the electric toroidal dipole $\mathbf{G}$ associated with ferroaxiality, and the electric toroidal quadrupoles are all forbidden by symmetry, so that the lowest-rank axial multipole available is $G_{xyz}$, the only axial octupole allowed in $T_h$, which belongs to the $A_{2g}$ representation of the parent $O_h$ point group \cite{hayami2018classification,yatsushiro2021multipole}. No other rank-three multipole is available. The electric octupole and the magnetic toroidal octupole are parity-odd and the magnetic octupole is time-reversal odd, so all three are excluded in a centrosymmetric nonmagnetic crystal. The octupole $G_{xyz}$ reverses its polarity under mirror reflection across planes containing the $\langle111\rangle$ direction [Fig. \ref{A2g}(a)].

The rank is also constrained directly by the measured direction dependence. The cross-circular intensity difference must be a pseudoscalar, invariant under the proper rotations of the point group and odd under the improper ones. The only direction available to build it from is the propagation direction $\hat{\mathbf{n}}$, which is a polar vector, so an axial multipole of rank $l$ enters contracted with $l$ factors of $\hat{\mathbf{n}}$. The leading term of the direction dependence therefore carries as many factors of $\hat{\mathbf{n}}$ as the rank of the leading multipole. For $l=3$ the contraction is $G_{xyz}\,n_x n_y n_z$ and it accounts for the measurements. Over the four \{111\} facets $n_x n_y n_z$ takes the signs $+$ $-$ $+$ $-$, which is the alternation of Table~\ref{t1}. A rank-one ferroaxial moment would instead give $\mathbf{G}\cdot\hat{\mathbf{n}}$ and no vector $\mathbf{G}$ can produce the observed $+\,-\,+\,-$ alternation, because $(111)$ and $(\bar1\bar11)$ on the one hand and $(\bar111)$ and $(1\bar11)$ on the other each sum to $2G_z$. The difference also vanishes for $\hat{\mathbf{n}}\parallel\langle001\rangle$, where the measured $|g_{\mathrm{ROA}}|$ is $0.03$, below the fitting uncertainty of $0.04$ (Table~\ref{t1}), and for $\hat{\mathbf{n}}\parallel\langle110\rangle$. In both of these directions the vanishing is exact rather than a leading-order result, because $T_h$ retains a mirror plane that contains the propagation direction, namely the $\{100\}$ planes normal to $[100]$ and $[010]$ for $\langle001\rangle$ and the $(001)$ plane for $\langle110\rangle$, so neither discriminates between multipoles of different rank. This octupolar symmetry has a simple real-space origin [Figs. \ref{A2g}(b) and \ref{A2g}(c)]. Each Fe site has $\bar{3}$ ($S_6$) site symmetry, so its FeS$_6$ cluster, together with the S$_2$ dimers bonded to each S, is rotated about the local $\langle111\rangle$ axis with respect to the crystal axes and thereby carries a ferroaxial (axial-vector) moment. Because pyrite is nonsymmorphic ($Pa\bar{3}$), the four Fe sublattices are related by $\langle100\rangle$ twofold operations combined with fractional translations. The resulting moments, directed along the four $\langle111\rangle$ axes, form a periodic all-in-all-out texture on the network of Fe tetrahedra, which realizes $G_{xyz}$. The experimentally observed sign alternation of the circular intensity difference between adjacent \{111\} faces, together with its strong suppression on the \{001\} surface, is therefore consistent with the presence of this electric toroidal octupolar symmetry.

\subsection{Exclusion of alternative mechanisms}

Spatial mapping shows that $g_{\mathrm{ROA}}$ is homogeneous within each facet and its sign is locked to the crystallographic orientation [Supplemental Material \cite{SM} (Sec. S6)], providing a facet-resolved, real-space visualization of the octupolar symmetry. Extrinsic origins, including residual strain, surface reconstruction, instrumental asymmetry, magnetic order, and optical anisotropy effects, are examined in the Supplemental Material \cite{SM} (Sec. S6) and found not to account for the observations. The same analysis further demonstrates that alternative mechanisms, including chiral-phonon effects and lower-rank axial multipolar symmetries, cannot account for the observed facet-dependent sign reversal, identifying $G_{xyz}$ as the lowest-rank symmetry-allowed origin.

The present effect is distinct from phonon splitting and other manifestations of phonon angular momentum \cite{zhu2018observation,Lujan2024,Juraschek2025} and from the $\Eg$ phonons predicted for cubic magnets with time-reversal-odd octupolar order \cite{hart2025phonon,sutcliffe2025pseudo}, whose signature is a lifting of the doublet degeneracy. Each of the two $\Eg$ components carries no net mechanical angular momentum at $\Gamma$ and along $\langle111\rangle$ inversion and time reversal make the $\mathbf{L}$ of $E^{+}$ and $E^{-}$ exactly opposite, so that $\mathbf{k}\cdot\mathbf{L}$ cancels between them (Fig.~\ref{FeS2_band}). The circular intensity difference is therefore not governed by chiral phonons but by the CAM $m=\pm1$ rendered inequivalent by the octupole. The effect thus provides a direct optical probe of a higher-rank axial multipole, a parity- and time-reversal-even symmetry of the host crystal that has no lower-rank counterpart.

\subsection{Role of resonance}
\label{sec:resonance}

The mirror that would interchange the $\Egn{+}$ and $\Egn{-}$ components, a reflection across a plane containing $\langle111\rangle$, is absent in $T_h$ (Sec.~\ref{sec:octupole}). Had it been retained, it would enforce $|\chi_{-}|=|\chi_{+}|$ and hence $U_{\mathrm{CC}}=0$ identically, so the inequivalence $|\chi_{-}|\neq|\chi_{+}|$ that renders $U_{\mathrm{CC}}$ finite is itself a consequence of $G_{xyz}$.
The octupolar symmetry is necessary but not sufficient. In the quasistatic (polarizability) limit the Raman tensor is real, so the two scattering channels are related by complex conjugation, $|\chi_{-}|=|\chi_{+}|$, and $U_{\mathrm{CC}}$ vanishes identically even when the octupole is present. A finite circular intensity difference therefore also requires an electronic resonance, which makes the Raman tensor complex and the two channels unequal~\cite{vogt1983complete}. The complexity originates in absorption rather than in broken time-reversal symmetry, which is why the Stokes and anti-Stokes differences share the same sign (Sec.~\ref{sec:facet}). The same requirement is long established for antisymmetric Raman scattering in molecules~\cite{barron2009molecular}. Because pyrite is opaque throughout the visible range, its Raman scattering is intrinsically resonant and this condition is naturally met here.
This distinguishes the present effect from the
helicity-dependent Raman responses of low-symmetry achiral crystals such
as ReS$_2$ and black phosphorus
\cite{zhang2017anomalous,han2022complex}, where the signal originates in the optical anisotropy of these layered crystals, which show linear birefringence and dichroism. The interference of these two can produce a chiroptical response even in a centrosymmetric crystal, but it requires a biaxial material and is therefore confined to the triclinic, monoclinic, and orthorhombic systems \cite{parrish2025differential}. Both comparison materials are of that kind, ReS$_2$ being triclinic and black phosphorus orthorhombic, whereas pyrite is cubic and optically isotropic, so that mechanism is absent here [Supplemental Material \cite{SM} (Sec. S6)].

The pronounced wavelength dependence of the ROA, including the sign reversal observed at 785 nm, further indicates an important role of the electronic intermediate states involved in the Raman process. Circular intensity differences are known to vary strongly with excitation energy in other materials as well~\cite{kusuno2026raman,zhang2022quantum}. Here the near-resonant weighting of the susceptibility components changes the sign as well as the magnitude. 
A phenomenological description of this resonant enhancement is given in the Supplemental Material \cite{SM} (Sec. S5).

The comparison between the calculated and the measured magnitudes deserves comment.
The symmetry-dictated content is reproduced, namely the $\Eg$ selectivity and the vanishing for the $\Tg$ triplets. The magnitudes are a separate matter and the mismatch at $1.60$~eV noted in Sec.~\ref{sec:calcspec} reflects the resonance profile rather than the symmetry analysis. Because
$U_{\mathrm{CC}}$ arises from interference between resonant denominators, $g_{\mathrm{ROA}}$
varies steeply with excitation energy. The measured value runs from $+0.85$ to $-0.08$ between
633 and 785~nm (Table~\ref{t1}), which are $0.38$~eV apart. In pyrite the intermediate states
are unusually sensitive to the structure, because the conduction-band minimum is built from the
antibonding $\sigma^{*}$ orbital of the S$_2$ dimer and its position is controlled primarily by
the S--S bond length \cite{eyert1998electronic,kolb2013ultrafast}. Two independent estimates of
that sensitivity, one from varying the internal parameter \cite{eyert1998electronic} and one
from the thermal excursion of the same bond \cite{kolb2013ultrafast}, give $0.10$ and
$0.13$~eV per $1\%$ of the bond length. The PBE structure used here overestimates that bond by
about $1\%$ [Supplemental Material \cite{SM} (Sec. S2)], so the calculated resonance structure
is expected to be displaced by $0.10$--$0.13$~eV and toward lower photon energy because the
edge lies too low \cite{eyert1998electronic}. That is the direction required for a calculated
difference that is still large at $1.60$~eV where the measured one has already decayed and it
is a displacement on the scale over which the computed cross-circular indicator changes by an
amount of order unity~\cite{watanabe2025dual}. A magnitude computed at a fixed photon energy is therefore not a quantitative prediction of the magnitude measured at the corresponding wavelength. The calculation nevertheless reproduces the reversal of sign with excitation energy, which symmetry alone does not fix. The octupolar assignment rests on the symmetry-dictated content, namely which mode
carries the effect and how its sign behaves under the facet and mirror operations, and is
unaffected by this limitation.

\section{Conclusions}

We have measured circularly polarized Raman scattering on the \{111\} and \{001\} facets of
pyrite FeS$_2$ at excitation wavelengths of 532, 633, and 785 nm. A circular intensity
difference appears exclusively for the doubly degenerate $\Eg$ phonon, reaching
$|g_{\mathrm{ROA}}| \simeq 0.85$ at 633 nm. Its sign alternates as $+$ $-$ $+$ $-$ across the
four neighboring \{111\} facets, is the same for Stokes and anti-Stokes scattering, and is
strongly suppressed on \{001\}. Symmetry analysis accounts for each of these observations. The
cross-circular intensity difference transforms as the antisymmetric part of the self-product of
the mode representation, which contains the octupolar $A_{2g}$ channel for the $\Eg$ doublet
but not for the $\Tg$ triplets, and it is odd under any mirror reflection whose plane contains
the propagation direction. First-principles resonant Raman calculations within the framework of
Ref.~\cite{watanabe2025dual}, evaluated here at the three experimental photon energies, reproduce
the $\Eg$ selectivity and the reversal of sign with excitation energy.

Our results establish phonon ROA as a tabletop, room-temperature probe of high-rank axial
multipolar symmetries, complementary to conventional probes that rely on macroscopic
polarization or magnetization, or on large-scale facilities~\cite{bhowal2021revealing}. Because
the response is resolved facet by facet, it further enables real-space imaging of the octupolar
symmetry. The underlying selection rule is not specific to the $\Eg$ phonon. It applies to any
degenerate excitation whose self-product contains the octupolar $A_{2g}$ representation in its
antisymmetric part and whose $m=\pm1$ sectors are resolved in the scattering geometry,
including other degenerate phonon branches and electronic excitations. It extends also to other crystal classes that permit the same octupole, such as the chiral class $23$, where a companion study reports cross-circular ROA in CoSi~\cite{CoSi2026}.

A sharper test is available in pyrite itself. Since the circular intensity difference is fixed
by where the electronic resonances lie relative to the photon energy, anything that moves those
states at constant structure type should move $g_{\mathrm{ROA}}$. Pressure does so without a
structural transition up to at least $55$~GPa~\cite{kleppe2004high}. The indirect gap
closes at $1.13(9)\times10^{-2}$~eV/GPa, measured to $28$~GPa~\cite{cervantes2002band}, and the resonance that enhances the
$\Ag$ mode, which is not the one probed here [Supplemental Material \cite{SM} (Sec. S5)],
moves downward by $0.32$~eV up to $51$~GPa~\cite{kleppe2004high}, comparable to the
$0.38$~eV across which $g_{\mathrm{ROA}}$ reverses sign here. If the states governing the
present response follow the same trend, the difference at 785~nm should grow under pressure.
The same measurements show that the $\Eg$ intensity falls relative to the other modes under
compression~\cite{kleppe2004high}, which sets the difficulty of the experiment.

\begin{acknowledgments}
T.S. was supported by JSPS KAKENHI (Grants No. JP19K21854 and No. JP26H02234), MEXT X-NICS (Grant No. JPJ011438), NINS OML Project (Grant No. OML012301), JST CREST (Grant No. JPMJCR24R5), and JST ERATO (Grant No. JPMJER2503). G.K. was supported by JST SPRING (Grant No. JPMJSP2180) and the Science Tokyo Support Program for Doctoral Students, funded by the Universities for International Research Excellence.
H.W. was supported by JSPS KAKENHI (Grants No. JP23K13058, No. JP24K00581, and No. JP25H02115) and JSR Corporation via JSR-UTokyo Collaboration Hub, CURIE.
R.O. was supported by JSPS KAKENHI (Grants No. JP26H00618 and No. JP26K17075). H.M. was supported by JSPS KAKENHI (Grant No. JP25K23349).
R.A. was supported by JSPS KAKENHI (Grants No. JP25H01246 and No. JP25H01252), JST CREST (Grant No. JPMJCR23O4), and RIKEN TRIP initiative (RIKEN Quantum, Advanced General Intelligence for Science Program, Many-body Electron Systems).
\end{acknowledgments}

\appendix

\section{Complex Raman tensor for the $\Eg$ phonons}
\label{app:tensors}

In the $T_h$ point group, the doubly degenerate $\Eg$ phonon at $\Gamma$ has two complex eigenvectors $\bm{u}_{\pm}$ that transform as the two one-dimensional complex irreducible representations under the threefold rotation $C_3^+$ about the $\langle 111\rangle$ axis \cite{watanabe2025dual},
\begin{equation}
C_3^{+}\bm{u}_{\pm} = e^{\mp 2\pi \mathrm{i}/3}\,\bm{u}_{\pm},
\qquad
\xi = e^{2\pi i/3}.
\label{eq:A1}
\end{equation}
so that $\bm{u}_{\pm}$ carries $m=\pm1$. We write $\Phi_{\pm}$ for the normal-coordinate amplitude
that multiplies $\bm{u}_{\mp}$, so that $\Phi_{\pm}\rightarrow\xi^{\pm1}\Phi_{\pm}$.
In the corresponding complex basis, the Raman susceptibility tensors associated with $\Phi_{+}$ and $\Phi_{-}$ can be written in diagonal form as~\cite{watanabe2025dual}
\begin{equation}
\hat{\chi}^{(+)}
=
\chi_{+}
\begin{pmatrix}
1 & 0 & 0 \\
0 & \xi^{-1} & 0 \\
0 & 0 & \xi
\end{pmatrix},
\quad
\hat{\chi}^{(-)}
=
\chi_{-}
\begin{pmatrix}
1 & 0 & 0 \\
0 & \xi & 0 \\
0 & 0 & \xi^{-1}
\end{pmatrix},
\label{eq:A2}
\end{equation}
where $\chi_{\pm} = (\hat{\chi}^{(\pm)})_{xx}$ denote the corresponding scalar Raman susceptibilities. In the notation of Ref.~\cite{watanabe2025dual}, $\hat{\chi}^{(\pm)}$ and $\Phi_{\pm}$ are $\hat{\chi}^{(1,2)}$ and $\Phi_{1,2}$. In Ref.~\cite{kusuno2026raman} they are $\hat{\chi}^{Q_{1,2}}$ and $Q_{1,2}$, with $\hat{\mathcal{R}}_{\pm}$ corresponding to $R_{1,2}$, and in both the index $1$ labels the LR channel.
The total Raman tensor for the doubly degenerate $\Eg$ mode is then given by
\begin{equation}
\hat{\alpha}
=
\Phi_{+} \hat{\chi}^{(+)}
+
\Phi_{-} \hat{\chi}^{(-)}.
\label{eq:A3}
\end{equation}

We write the optical fields with the time dependence $e^{-\mathrm{i}\omega t}$, so that a unit polarization vector $\bm{e}$ carries the photon spin $\mathrm{Im}(\bm{e}^{*}\times\bm{e})\cdot\hat{\mathbf{n}}$ in units of $\hbar$ along the incident propagation direction $\hat{\mathbf{n}}$, which serves as the reference for both beams. Right and left circular are named by the sign of that spin, positive and negative, respectively, as in the convention in which a right-circularly polarized photon carries positive angular momentum along its propagation direction~\cite{yariv2007}.
For circularly polarized light propagating along the $[111]$ direction, we choose the polarization vectors
\begin{equation}
\bm{e}_{\mathrm{L}}=\frac{1}{\sqrt{3}}
\begin{pmatrix}
1\\
\xi^{-1}\\
\xi
\end{pmatrix},
\qquad
\bm{e}_{\mathrm{R}}=\frac{1}{\sqrt{3}}
\begin{pmatrix}
1\\
\xi\\
\xi^{-1}
\end{pmatrix}.
\label{eq:A4}
\end{equation}
These carry photon spin $-\hbar$ and $+\hbar$ along $[111]$, respectively, and are eigenvectors of $C_3^+$ with eigenvalues $\xi$ and $\xi^{-1}$, so that $m=-1$ and $m=+1$, respectively. With this convention $m$ coincides with the photon spin in units of $\hbar$.
For a Stokes process, conservation of crystal angular momentum about $\langle111\rangle$ gives
\begin{equation}
m \equiv \sigma_{\mathrm{i}} - \sigma_{\mathrm{s}} = \mp 2 \equiv \pm 1
\ \ (\mathrm{mod}\ 3),
\label{eq:A4b}
\end{equation}
the upper sign corresponding to the LR configuration, since a threefold axis conserves angular
momentum only modulo three. Here $\sigma_{\mathrm{i}}$ and $\sigma_{\mathrm{s}}$ are the CAM
labels of the incident and scattered polarization vectors and $m$ is the CAM of the phonon
created in the Stokes process. Upon quantization, because the two complex $\Eg$ eigenvectors are
conjugate partners and the lattice-displacement operator is Hermitian, the normal-coordinate
operator associated with $\Phi_{+}$ contains the creation operator of the $m=+1$ phonon, whereas
that associated with $\Phi_{-}$ contains the creation operator of the $m=-1$ phonon. Thus the LR
configuration couples to $\Phi_{+}$ and creates the $m=+1$ phonon, while the RL configuration
couples to $\Phi_{-}$ and creates the $m=-1$ phonon.
Accordingly, the index of $\Phi_{\pm}$, $\chi_{\pm}$, $\hat{\chi}^{(\pm)}$, and
$\hat{\mathcal{R}}_{\pm}$ labels the phonon created in the Stokes process, as does the superscript
of $\Egn{\pm}$.
The Raman intensity is proportional to
\begin{equation}
I \propto \left| \bm{e}_s^\dagger \hat{\alpha} \bm{e}_i \right|^2 .
\label{eq:A5}
\end{equation}
Evaluating Eq.~(\ref{eq:A5}) for the cross-circular configurations yields
\begin{equation}
I^{[111]}_{\mathrm{LR}}
=
\left| \bm{e}_{\mathrm{R}}^\dagger \hat{\alpha} \bm{e}_{\mathrm{L}} \right|^2
=
|\chi_{+} \Phi_{+}|^2,
\label{eq:A6}
\end{equation}
and
\begin{equation}
I^{[111]}_{\mathrm{RL}}
=
\left| \bm{e}_{\mathrm{L}}^\dagger \hat{\alpha} \bm{e}_{\mathrm{R}} \right|^2
=
|\chi_{-} \Phi_{-}|^2.
\label{eq:A7}
\end{equation}
Thus the LR and RL configurations couple to $\Phi_{+}$ and $\Phi_{-}$, respectively, and in a Stokes process create phonons with CAM $m=+1$ and $m=-1$, respectively.

The circular intensity difference is therefore given by
\begin{equation}
U^{[111]}_{\mathrm{CC}}
=
I_{\mathrm{LR}}-I_{\mathrm{RL}}
=
|\chi_{+} \Phi_{+}|^2-|\chi_{-} \Phi_{-}|^2 .
\label{eq:A8}
\end{equation}

Under the mirror operations relating neighboring \{111\} faces, the two scattering channels are interchanged as a whole, $(\chi_{+},\Phi_{+})\leftrightarrow(\chi_{-},\Phi_{-})$, so that $U_{\mathrm{CC}}^{[\bar111]}=|\chi_{-}\Phi_{-}|^2-|\chi_{+}\Phi_{+}|^2=-U_{\mathrm{CC}}^{[111]}$. The sign of $U_{\mathrm{CC}}$ therefore alternates between adjacent \{111\} faces.

Equation~(\ref{eq:A8}) is the tensor-level counterpart of the
representation-theoretic selection rule of Sec.~\ref{sec:rules}. A
nonzero $U_{\mathrm{CC}}$ requires $|\chi_{+}\Phi_{+}|\neq|\chi_{-}\Phi_{-}|$,
an inequivalence that the octupole can produce only through the $A_{2g}$
channel, which is present in $\Eg\otimes\Eg$ and absent in
$T_{2g}\otimes T_{2g}$.

The diagonal representation in Eq.~(\ref{eq:A2}) is equivalent to the cyclic tensor form used in the Cartesian basis. The two are related by a basis transformation to the complex $C_3$ eigenbasis. We note that the use of a complex Raman tensor to describe the response of degenerate vibrational modes to circularly polarized light is well established in the theory of Raman optical activity \cite{barron2009molecular}.

\clearpage


%

\bibliographystyle{apsrev4-2-title}
\bibliography{refs}

\end{document}

%% file: revpackage.tex
\usepackage{graphicx}

\usepackage{bm}
\usepackage{xcolor}
    
\usepackage{amsmath, amssymb, amsfonts}

\usepackage[colorlinks=true,allcolors=blue]{hyperref}

\usepackage{siunitx} 
\NewCommandCopy\sqty\qty 
\NewDocumentCommand\Sqty{smm}{%
    \IfBooleanTF{#1}%
    {\sqty[quantity-product={}]{#2}{#3}}%
    {\sqty[quantity-product=~]{#2}{#3}}%
}
\usepackage[italicdiff]{physics} 
\usepackage[version=4]{mhchem} 

\usepackage{cleveref}
\crefname{equation}{Eq.}{Eqs.} 

\newcommand{\Ag}{A_g}

\newcommand{\Eg}{E_g}

\newcommand{\Egn}[1]{\Eg^{#1}}

\newcommand{\Tg}{T_g}

\newcommand{\IRL}{I_{\mathrm{RL}}}
\newcommand{\ILR}{I_{\mathrm{LR}}}

\NewDocumentCommand\qtxt{sm}{%
    \IfBooleanTF{#1}%
    {~\textrm{#2}}%
    {\quad\textrm{#2}}%
}

\NewDocumentCommand{\subref}{mm}{\hyperref[#1]{\ref{#1}(#2)}}
\graphicspath{%
{../fig/pdf/}%
}

\usepackage{braket}